\documentclass[prd,twocolumn,groupedaddress,showpacs,nofootinbib]{revtex4}
\usepackage{graphicx}
\usepackage{dcolumn}
\usepackage{amssymb}
\usepackage{mathrsfs}
\usepackage{amsmath}
\usepackage{epsfig}
\usepackage[dvips]{color}
\usepackage{hhline}

\begin{document}

\title{Weinberg dimension-5 operator by vector-like lepton doublets }

\author{Pei-Hong Gu}

\email{phgu@seu.edu.cn}

\affiliation{School of Physics, Jiulonghu Campus, Southeast University, Nanjing 211189, China}

\begin{abstract}

It is well known that a Weinberg dimension-5 operator for small neutrino masses can be realized at tree level in three types of renormalizable models: (i) the type-I seesaw mediated by fermion singlets, (ii) the type-II seesaw mediated by Higgs triplets, (iii) the type-III seesaw mediated by fermion triplets. We here point out such operator can be also induced at tree level by vector-like lepton doublets in association with unusual fermion singlets, Higgs triplets or fermion triplets. If these unusual fermion singlets, Higgs triplets or fermion triplets are heavy enough, their decays can generate a lepton asymmetry to explain the cosmic baryon asymmetry, meanwhile, the vector-like lepton doublets can lead to a novel inverse or linear seesaw with rich observable phenomena. We further specify our scenario can be naturally embedded into a grand unification theory without the conventional type-I, type-II or type-III seesaw.

\end{abstract}


\maketitle

The discovery of neutrino oscillations indicates that three flavors of neutrinos should be massive and mixing \cite{pdg2018}. Within the context of the standard model (SM), we can consider a Weinberg dimension-5 operator \cite{weinberg1979},
\begin{eqnarray}
\label{weinberg}
\mathcal{L}&\supset& -\frac{1 }{2\Lambda} \bar{l}_L^{}\phi \phi^T_{} l_L^{c} + \textrm{H.c.}\,,
\end{eqnarray}
to generate the required neutrino masses,
\begin{eqnarray}
\mathcal{L}&\supset& -\frac{1 }{2} m_\nu^{} \bar{\nu}_L^{}\nu_L^{c} + \textrm{H.c.}~~\textrm{with}~~m_\nu^{}=\frac{\langle\phi\rangle^2_{} }{\Lambda}\,.
\end{eqnarray}
In Eq. (\ref{weinberg}), $l_L^{}$ and $\phi$ are the lepton and Higgs doublets, 
\begin{eqnarray}
\begin{array}{c}l_L^{}(1,2,-\frac{1}{2})\end{array}\,,~~\begin{array}{c}\phi(1,2,-\frac{1}{2})\end{array}\,.
\end{eqnarray}
Here and thereafter the brackets following the fields describe the transformations under the $SU(3)_c^{} \times SU(2)^{}_{L}\times U(1)_Y^{}$ gauge groups.

\begin{figure*}
\vspace{6.5cm} \epsfig{file=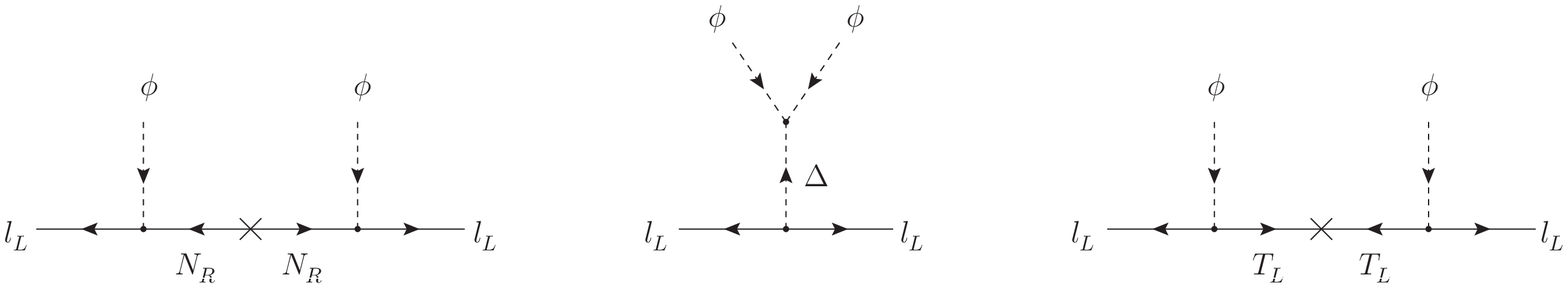, bbllx=10.5cm, bblly=6.0cm,
bburx=20.5cm, bbury=16cm, width=7cm, height=7cm, angle=0,
clip=0} \vspace{-10cm} \caption{\label{numass123} The Weinberg dimension-5 operator induced by the tyep-I, type-II and type-III seesaw at tree level.}
\end{figure*}

It has been well known that the Weinberg dimension-5 operator (\ref{weinberg}) can be realized at tree level in three types of renormalizable models: (i) the type-I seesaw mediated by fermion singlets \cite{minkowski1977,yanagida1979,grs1979,ms1980}, (ii) the type-II seesaw mediated by Higgs triplets \cite{mw1980,sv1980,cl1980,lsw1981,ms1981}, (iii) the type-III seesaw mediated by fermion triplets \cite{flhj1989,ma1998}, i.e.  
\begin{eqnarray}
\label{lag1}
\mathcal{L}_{\textrm{I}}^{}&\supset& - \frac{1}{2}M_N^{}\left( \bar{N}_R^{c} N_R^{} +\textrm{H.c.}\right)-y_N^{} \bar{l}_L^{} \phi N_R^{} + \textrm{H.c.}\nonumber\\
&&\textrm{with}~~N_R^{}(1,1,0)\,,\\
[2mm]
\label{lag2}
\mathcal{L}_{\textrm{II}}^{}&\supset& -M_\Delta^2 \textrm{Tr}\left(\Delta^\dagger_{}\Delta\right) - \mu_\Delta^{}\left(\phi^T_{} i \tau_2^{} \Delta \phi +\textrm{H.c.}\right)\nonumber\\
&&- \frac{1}{2}f_\Delta^{} \bar{l}_L^c i\tau_2^{} \Delta l_L^{} + \textrm{H.c.}~~\textrm{with}~~\Delta(1,3,+1) \,,\\
[2mm]
\label{lag3}
\mathcal{L}_{\textrm{III}}^{}&\supset& - \frac{1}{2}M_T^{} \left[\textrm{Tr}\left(\bar{T}_L^{c} i\tau_2^{} T_L^{} i \tau_2^{}\right)+\textrm{H.c.}\right] -y_T^{} \bar{l}_L^{c} i\tau_2^{} T_L^{} \tilde{\phi} \nonumber\\
&&+ \textrm{H.c.}~~\textrm{with}~~T_L^{}(1,3,0) \,.
\end{eqnarray}
As shown in Fig. \ref{numass123}, we can integrate out the right-handed neutrino singlets $N_R^{}$, the Higgs triplets $\Delta$ and the lepton triplets $T_L^{}$ from the type-I, II and III seesaw models. Then the effective cutoff in Eq. (\ref{weinberg}) should be 
\begin{eqnarray}
\frac{1}{\Lambda_{\textrm{I}}^{}} &=&-y_N^{}\frac{1}{M_N^{}}y_N^T\,,~
\frac{1}{\Lambda_{\textrm{II}}^{}} =-f_\Delta^{\dagger}\frac{\mu_\Delta^{}}{M_\Delta^{2}}\,,\nonumber\\
\frac{1}{\Lambda_{\textrm{III}}^{}} &=&-y_T^{\ast}\frac{1}{M_T^{}}y_T^\dagger\,.
\end{eqnarray}

 \begin{figure*}
\vspace{7cm} \epsfig{file=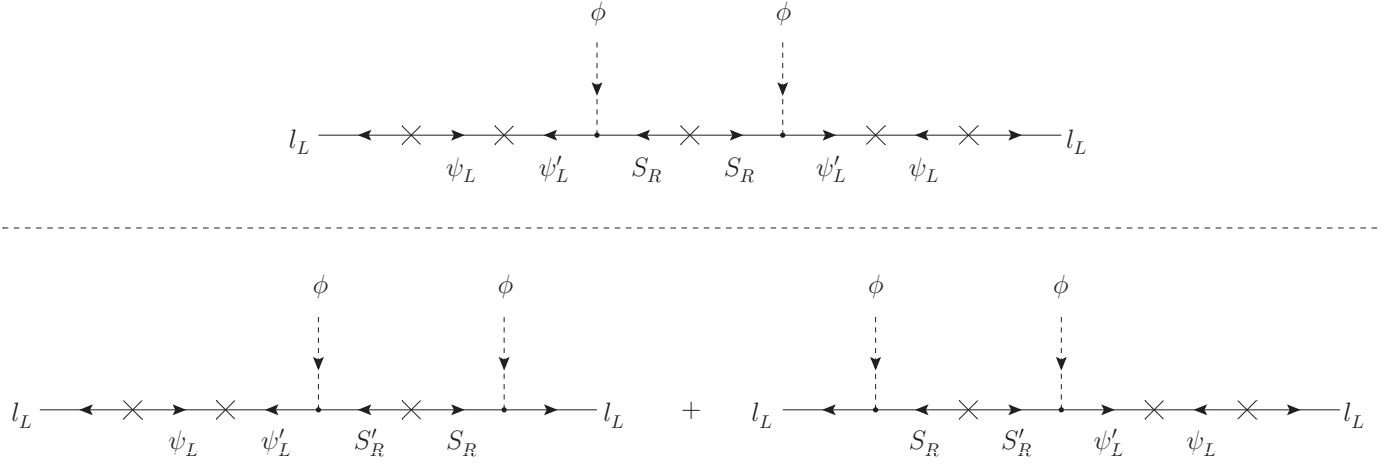, bbllx=10.5cm, bblly=6.0cm,
bburx=20.5cm, bbury=16cm, width=7cm, height=7cm, angle=0,
clip=0} \vspace{-7cm} \caption{\label{numass45}  The Weinberg dimension-5 operator induced by the vector-like lepton doublets and the fermion singlets at tree level. }
\end{figure*}

\begin{figure*}
\vspace{6.5cm} \epsfig{file=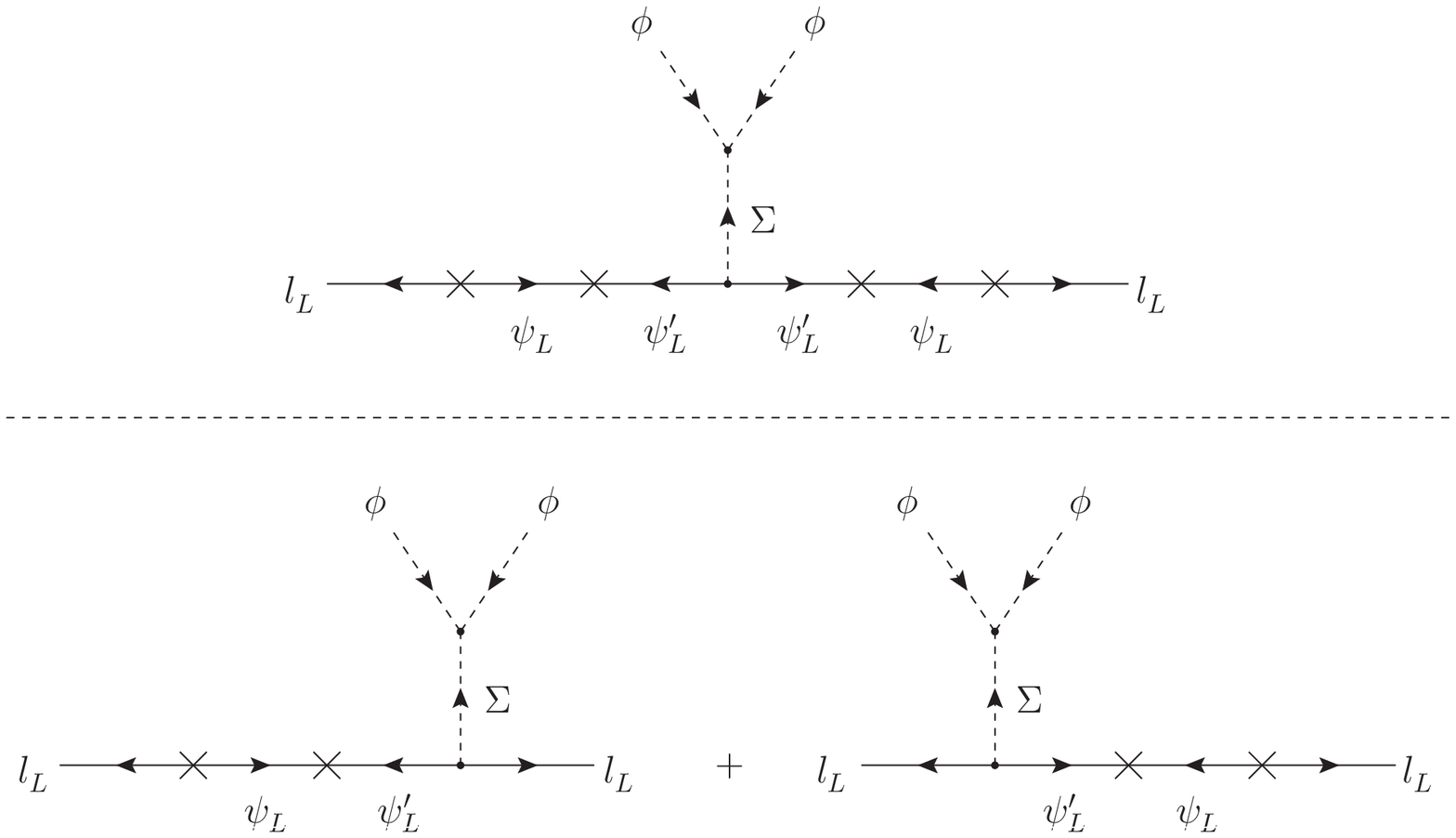, bbllx=7.5cm, bblly=6.0cm,
bburx=17.5cm, bbury=16cm, width=7cm, height=7cm, angle=0,
clip=0} \vspace{-5.5cm} \caption{\label{numass67}  The Weinberg dimension-5 operator induced by the vector-like lepton doublets and the Higgs triplets at tree level.}
\end{figure*}

\begin{figure*}
\vspace{6.5cm} \epsfig{file=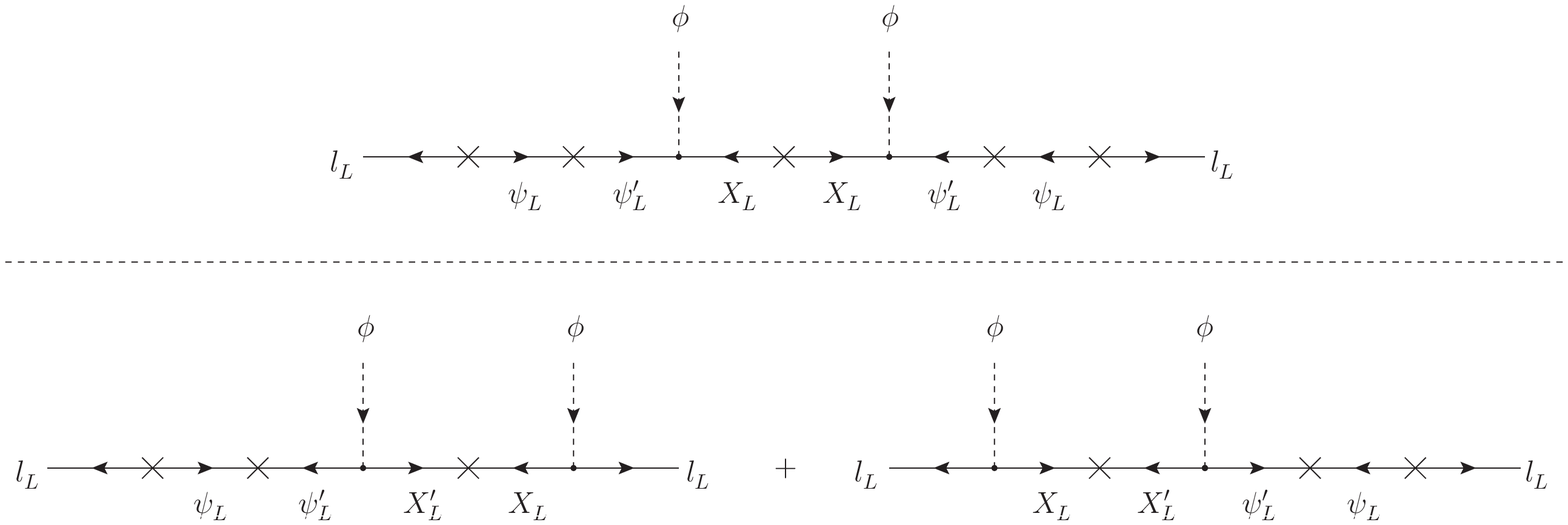, bbllx=10.5cm, bblly=6.0cm,
bburx=20.5cm, bbury=16cm, width=7cm, height=7cm, angle=0,
clip=0} \vspace{-7cm} \caption{\label{numass89}  The Weinberg dimension-5 operator induced by the vector-like lepton doublets and the fermion triplets at tree level. }
\end{figure*}

 \begin{figure*}
\vspace{7.5cm} \epsfig{file=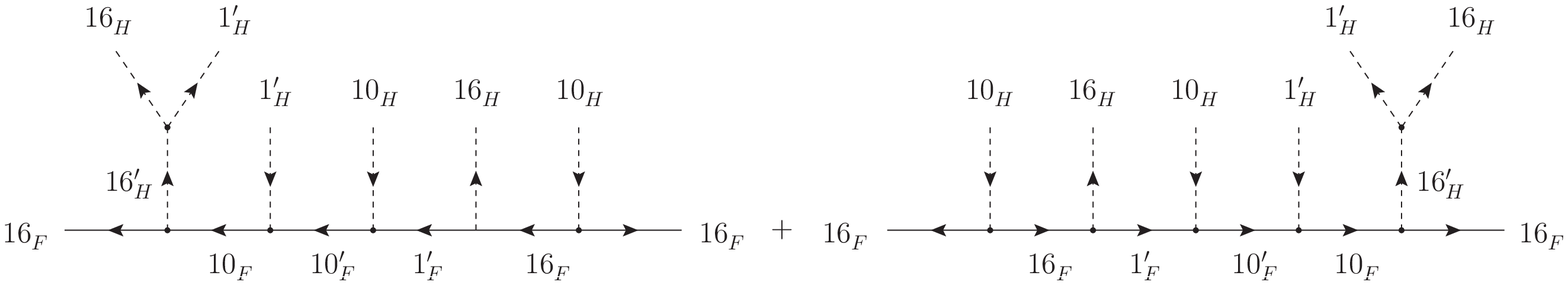, bbllx=10.5cm, bblly=6.0cm,
bburx=20.5cm, bbury=16cm, width=7cm, height=7cm, angle=0,
clip=0} \vspace{-9.5cm} \caption{\label{numassgut}  The Weinberg dimension-5 operator from an $SO(10)$ model where the vector-like lepton doublets are from the $(10_F^{},10'^{}_F)$ representations while the fermion singlets are from the $(1'^{}_F,16^{}_F)$ representations. The $1'^{}_H$ Higgs scalar can obtain an induced VEV at the TeV scale after a PQ symmetry is spontaneously broken. The $16'^{}_H$ Higgs scalar then can obtain its VEV below the TeV scale after the $16^{}_H$ Higgs scalar spontaneously breaks the left-right symmetry. }
\end{figure*}

In the following we shall show the other renormalizable ways to realize the Weinberg dimension-5 operator (\ref{weinberg}) at tree level. Specifically, we introduce two types of fermion doublets with opposite hypercharges,
\begin{eqnarray}
\begin{array}{c}\psi_L^{}(1,2,+\frac{1}{2})\end{array}\,,~~\begin{array}{c}\psi'^{}_L(1,2,-\frac{1}{2})\end{array}\,.
\end{eqnarray}
We then construct the models in association with fermion singlets, 
\begin{subequations}
\begin{eqnarray}
\label{lag4a}
\mathcal{L}_{\textrm{IVa}}^{}&\supset& -\frac{1}{2}M_S^{}\left( \bar{S}_R^c S_R^{} +\textrm{H.c.}\right) -M_\psi^{} \left(\bar{\psi}^{c}_L i\tau_2^{} \psi'^{}_L +\textrm{H.c.}\right)   \nonumber\\
&& - m_D^{} \bar{l}_L^c i\tau_2^{}  \psi_L^{} - f'^{}_S \bar{\psi}'^{}_L\phi S_R^{} +\textrm{H.c.}\nonumber\\
&&\textrm{with}~~ \begin{array}{c}S_R^{}(1,1,0)\,, \end{array}\\
[2mm]
\label{lag4b}
\mathcal{L}_{\textrm{IVb}}^{}&\supset& -M_S^{}\left(\bar{S}_R^c S'^{}_R+\textrm{H.c.}\right)   -M_\psi^{} \left(\bar{\psi}^{c}_L i\tau_2^{} \psi'^{}_L +\textrm{H.c.}\right)  \nonumber\\
&&- m_D^{} \bar{l}^c_L i\tau_2^{} \psi_L^{} - f^{}_S \bar{l}^{}_L\phi S_R^{}  - f'^{}_S \bar{\psi}'^{}_L\phi S'^{}_R  +\textrm{H.c.}\nonumber\\
&&\textrm{with}~~\begin{array}{c}S_R^{}(1,1,0)\,,\end{array} \begin{array}{c}S'^{}_R(1,1,0)\,,\end{array}\end{eqnarray}
\end{subequations}
the models in association with Higgs triplets,
\begin{subequations}
\begin{eqnarray}
\label{lag5a}
\mathcal{L}_{\textrm{Va}}^{}&\supset& -M_\Sigma^2 \textrm{Tr}\left(\Sigma^\dagger_{}\Sigma\right)    - \mu_\Sigma^{} \left(\phi^T_{} i \tau_2^{} \Sigma \phi +\textrm{H.c.}\right)  \nonumber\\
&&-M_\psi^{} \left(\bar{\psi}^{c}_L i\tau_2^{} \psi'^{}_L +\textrm{H.c.}\right) - m_D^{} \bar{l}_L^c i\tau_2^{}  \psi_L^{} \nonumber\\
&& - \frac{1}{2}f_\Sigma^{} \bar{\psi}'^c_L i\tau_2^{} \Sigma \psi'^{}_L +\textrm{H.c.}~~\textrm{with}~~\begin{array}{c}\Sigma(1,3,+1)\,, \end{array}\nonumber\\
&&\\
[2mm]
\label{lag5b}
\mathcal{L}_{\textrm{Vb}}^{}&\supset& -M_\Sigma^2 \textrm{Tr}\left(\Sigma^\dagger_{}\Sigma\right)    - \mu_\Sigma^{} \left(\phi^T_{} i \tau_2^{} \Sigma \phi +\textrm{H.c.}\right)  \nonumber\\
&&-M_\psi^{} \left(\bar{\psi}^{c}_L i\tau_2^{} \psi'^{}_L +\textrm{H.c.}\right)- m_D^{} \bar{l}^c_L i\tau_2^{} \psi_L^{}  \nonumber\\
&&- f_\Sigma^{} \bar{\psi}'^c_L i\tau_2^{} \Sigma l^{}_L  +\textrm{H.c.}~~\textrm{with}~~ \begin{array}{c}\Sigma(1,3,+1)\,,\end{array}\nonumber\\
&&
\end{eqnarray}
\end{subequations}
as well as the models in association with fermion triplets, 
\begin{subequations}
\begin{eqnarray}
\label{lag6a}
\mathcal{L}_{\textrm{VIa}}^{}&\supset& -\frac{1}{2}M_X^{}\left[ \textrm{Tr}\left(\bar{X}^{c}_L i\tau_2^{} X^{}_L i\tau_2^{}\right) +\textrm{H.c.}\right] \nonumber\\
&&-M_\psi^{} \left(\bar{\psi}^{c}_L i\tau_2^{} \psi'^{}_L +\textrm{H.c.}\right)   - m_D^{} \bar{l}_L^c i\tau_2^{}  \psi_L^{} \nonumber\\
&&- f'^{}_X \bar{\psi}'^{c}_L \i \tau_2^{} X_L^{} \phi +\textrm{H.c.}~~\textrm{with}~~ \begin{array}{c}X^{}_L(1,3,0)\,, \end{array}\nonumber\\
&&\\
[2mm]
\label{lag6b}
\mathcal{L}_{\textrm{VIb}}^{}&\supset& - M_X^{}\left[ \textrm{Tr}\left(\bar{X}^{c}_L i\tau_2^{} X'^{}_L i\tau_2^{}\right) +\textrm{H.c.}\right] \nonumber\\
&&-M_\psi^{} \left(\bar{\psi}^{c}_L i\tau_2^{} \psi'^{}_L +\textrm{H.c.}\right)   - m_D^{} \bar{l}_L^c i\tau_2^{}  \psi_L^{} \nonumber\\
&&- f^{}_X \bar{l}^{c}_L i\tau_2^{} X_L^{} \phi   - f'^{}_X \bar{\psi}'^{c}_L i\tau_2^{} X'^{}_L \phi   +\textrm{H.c.}\nonumber\\
&&\textrm{with}~~ \begin{array}{c}X^{}_L(1,3,0)\,,\end{array}~~ \begin{array}{c}X'^{}_L(1,3,0)\,.\end{array} \end{eqnarray}
\end{subequations}
Here we have imposed a global symmetry of lepton number, which is softly broken only by the $m_D^{}$ terms, in order to forbid the other mass and Yukawa terms involving the non-SM fields. For this purpose, $(\psi_L^{},\psi'^{}_L,S^{}_R)$ carry the lepton numbers $L=(0,0,0)$ for the model (\ref{lag4a}), $(\psi_L^{},\psi'^{}_L,S^{}_R,S'^{}_R)$ carry $L=(+1,-1,+1,-1)$ for the model (\ref{lag4b}), $(\psi_L^{},\psi'^{}_L,\Sigma)$ carry $L=(0,0,0)$ for the model (\ref{lag5a}) or $L=(+1,-1,0)$ for the model (\ref{lag5b}), $(\psi^{}_L, \psi'^{}_L, X^{}_L)$ carry $L=(0,0,0)$ for the model (\ref{lag6a}), $(\psi^{}_L, \psi'^{}_L, X^{}_L,X'^{}_L)$ carry $L=(+1,-1,-1,+1)$ for the model (\ref{lag6b}). By integrating out the non-SM fields from Eqs. (\ref{lag4a}-\ref{lag6b}), the effective cutoff in Eq. (\ref{weinberg}) can be respectively given by 
\begin{subequations} 
 \begin{eqnarray}
\!\!\! \!\!\!\!\!\! \!\!\!\! \!\!\!\!\frac{1}{\Lambda_{\textrm{IVa}}^{}} \!\!&=&\!\!m_D^{\ast} \frac{1}{M_\psi^{}}f'^{}_S \frac{1}{M_S^{}} f'^{T}_S \frac{1}{M_\psi^{}}m_D^{\dagger} \,,\\
[2mm]
\!\!\!\!\! \!\!\!\! \!\!\!\! \!\!\!\!\frac{1}{\Lambda_{\textrm{IVb}}^{}} \!\!&=&\!\!-f_S^{}\frac{1}{M_S^{}}f'^T_S \frac{1}{M_\psi^{}} m_D^{\dagger}- m_D^{\ast}\frac{1}{M_\psi^{}}f'^{}_S\frac{1}{M_S^{}}f^T_S \,,
\end{eqnarray}
\end{subequations}
\begin{subequations}
\begin{eqnarray}
\!\!\!\!\!\! \!\!\!\! \!\!\!\! \!\!\!\! \!\!\!\! \!\!\!\!\frac{1}{\Lambda_{\textrm{Va}}^{}} \!\!&=&\!\!m_D^{\ast} \frac{1}{M_\psi^{}}f_\Sigma^{\dagger}\frac{1}{M_\psi^{}}m_D^{\dagger} \frac{\mu_\Sigma^{}}{M_\Sigma^{2}}\,,\\
[2mm]
\!\!\!\!\!\! \!\!\!\! \!\!\!\! \!\!\!\! \!\!\!\! \!\!\!\!\frac{1}{\Lambda_{\textrm{Vb}}^{}} \!\!&=&\!\!-\left(f_\Sigma^{\dagger}\frac{1}{M_\psi^{}}m_D^{\dagger}+m_D^{\ast}\frac{1}{M_\psi^{}}f_\Sigma^{\ast} \right)\frac{\mu_\Sigma^{}}{M_\Sigma^{2}}\,,
\end{eqnarray}
\end{subequations}
\begin{subequations}
\begin{eqnarray}
\!\!\!\!\!\!\! \!\!\!\!\frac{1}{\Lambda_{\textrm{VIa}}^{}} \!\!&=&\!\!m_D^{\ast} \frac{1}{M_\psi^{}}f'^{\ast}_X \frac{1}{M_X^{}} f'^{\dagger}_X \frac{1}{M_\psi^{}}m_D^{\dagger}\,,\\
[2mm]
\!\!\!\!\!\!\! \!\!\!\!\frac{1}{\Lambda_{\textrm{VIb}}^{}} \!\!&=&\!\!-f_X^{\ast}\frac{1}{M_X^{}}f'^\dagger_X \frac{1}{M_\psi^{}} m_D^{\dagger}- m_D^{\ast}\frac{1}{M_\psi^{}}f'^{\ast}_X\frac{1}{M_X^{}}f^\dagger_X\,.
\end{eqnarray}
\end{subequations}
The relevant diagrams are shown in Fig. \ref{numass45}, Fig. \ref{numass67} and Fig. \ref{numass89}.

If the vector-like lepton doublets $(\psi^{}_L,\psi'^{}_L)$ are heavy enough, they can be integrated out before the fermion singlets $(S^{}_R,S'^{}_R)$, the Higgs triplets $\Sigma$ or the fermion triplets $(X^{}_L,X'^{}_L)$. This means the models (\ref{lag4a}-\ref{lag6b}) can give nothing but the type-I, type-II or type-III seesaw. Alternatively, the fermion singlets $(S_R^{},S'^{}_R)$, the Higgs triplets $\Sigma$ and the fermion triplets $(X_L^{}, X'^{}_L)$ could be much heavier than the vector-like lepton doublets $(\psi^{}_L,\psi'^{}_L)$. In consequence, we could first integrate out these fermion singlets, Higgs triplets or fermion triplets and then study the phenomena from the vector-like lepton doublets. For example, the models (\ref{lag4a}), (\ref{lag5a}) and (\ref{lag6a}) indeed can lead to an inverse seesaw \cite{mv1986}. In this inverse seesaw scenario, the charged and neutral components of the vector-like lepton doublets $(\psi^{}_L,\psi'^{}_L)$ can respectively mix with the SM charged leptons and neutrinos up to the bounds allowed by the precision measurements \cite{abp2008}. Such significant mixings can be also induced by the $m_D^{}$ term in the models (\ref{lag4b}), (\ref{lag5b}) and (\ref{lag6b}), which now have a few features of the linear seesaw \cite{alsv1995,barr2004,mrv2005}. We hence could expect the models (\ref{lag4a}-\ref{lag6b}) to be verified at the LHC and other colliders, similar to the inverse seesaw with vector-like lepton triplets \cite{ma2009}.

It is easy to see our models can be embedded into an $SU(5)$ grand unification theory (GUT) by placing the vector-like lepton doublets in the 5-dimensional representations. We further consider the $SO(10)$ GUT. In Fig. \ref{numassgut}, we indicate the model (\ref{lag4b}) can originate from an $SO(10)$ GUT where the vector-like lepton doublets are from the $(10_F^{},10'^{}_F)$ representations \cite{ma2018} while the fermion singlets are from the $(1'^{}_F,16^{}_F)$ representations. In this model we have imposed a global $U(1)_{PQ}^{}$ symmetry under which the $16^{}_H$ field carries a charge $-1$, the $1'^{}_H$, $1'^{}_F$ and $10'^{}_F$ fields carry a charge $+1$, while the $16'^{}_H$ and $10^{}_F$ carry a charge $+2$. Consequently, the trilinear coupling $ \overline{10}^{}_H16^{}_H 16^{}_H$ should be absent and hence the $[SU(2)_L^{}]$-singlet components of the $16_H^{}$ Higgs scalar cannot obtain a vacuum expectation value (VEV). The $1'^{}_H$ Higgs scalar has a quartic coupling with another $1^{}_H$ Higgs scalar, i.e. $\bar{1}'^{}_H 1^3_H$. After the $1^{}_H$ Higgs scalar develops a VEV around $10^{10-12}_{}\,\textrm{GeV}$, the $1'^{}_H$ Higgs scalar can be expected to obtain a VEV around $10^{3-4}_{}\,\textrm{GeV}$ if its mass term is below the GUT scale. Since the $(10^{}_F, 10'^{}_F)$ fields contains the color-triplet fermions, the $U(1)_{PQ}^{}$ symmetry should be the Peccei-Quinn (PQ) symmetry \cite{pq1977} for a KSVZ invisible axion \cite{kim1979,svz1980}. We would like to emphasize this GUT does not result in the conventional type-I and type-II seesaw. Actually the usual right-handed neutrinos from the $16_F^{}$ representations now construct the fermion singlets with the $1'^{}_F$ representations. Similarly, the other models (\ref{lag4a}) and (\ref{lag5a}-\ref{lag6b}) can be also embedded into an $SO(10)$ context such as the above GUT for the model (\ref{lag4b}).

\begin{figure*}
\vspace{6.5cm} \epsfig{file=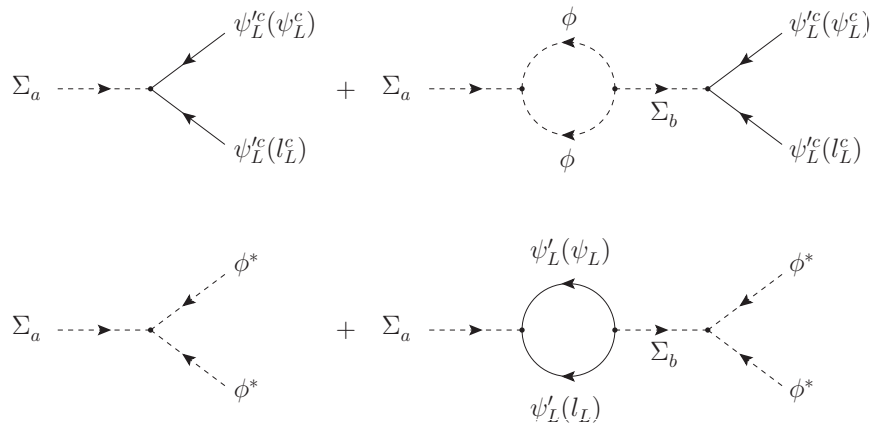, bbllx=5.5cm, bblly=6.0cm,
bburx=15.5cm, bbury=16cm, width=7cm, height=7cm, angle=0,
clip=0} \vspace{-7.5cm} \caption{\label{sdecay}  The heavy Higgs triplet decays at one-loop order. }
\end{figure*}

We then simply explain how to produce the cosmic baryon asymmetry in our models. For example, we consider the models (\ref{lag5a}) and (\ref{lag5b}). If and only if the kinematics is allowed, the Higgs triplets $\Sigma$ can have two decay modes as below,
 \begin{eqnarray}
 \label{cpvdecay}
\Sigma \rightarrow \psi'^c_L + \psi'^{c}_L~~\textrm{or}~~\psi^c_L+l^c_L\,,~~\Sigma \rightarrow \phi^\ast_{}+\phi^\ast_{}\,.
\end{eqnarray}
As long as the CP is not conserved, we can expect a CP asymmetry in the above decays, 
\begin{eqnarray}
\label{cpa}
\varepsilon_a^{}&=&2\frac{\Gamma (\Sigma_a^{} \rightarrow \psi'^c_L + \psi'^{c}_L) -\Gamma (\Sigma^\ast_{a} \rightarrow \psi'^{}_L + \psi'^{}_L)  }{\Gamma (\Sigma_a^{} \rightarrow \psi'^c_L + \psi'^{c}_L)+\Gamma(\Sigma_a^{} \rightarrow \phi^\ast_{}+\phi^\ast_{})}\nonumber\\
&=&2\frac{ \Gamma (\Sigma^\ast_{a} \rightarrow \phi + \phi) - \Gamma (\Sigma_a^{} \rightarrow \phi^\ast_{}+ \phi^\ast_{}) }{\Gamma (\Sigma_a^{} \rightarrow \psi'^c_L + \psi'^{c}_L)+\Gamma(\Sigma_a^{} \rightarrow \phi^\ast_{}+\phi^\ast_{})}\neq 0 ~~\textrm{or}\nonumber\\
[2mm]
\varepsilon_a^{}&=&2\frac{\Gamma (\Sigma_a^{} \rightarrow \psi^c_L + l^{c}_L) -\Gamma (\Sigma^\ast_{a} \rightarrow \psi^{}_L + l^{}_L)  }{\Gamma (\Sigma_a^{} \rightarrow \psi^c_L + l^{c}_L)+\Gamma(\Sigma_a^{} \rightarrow \phi^\ast_{}+\phi^\ast_{})}\nonumber\\
&=&2\frac{ \Gamma (\Sigma^\ast_{a} \rightarrow \phi + \phi) - \Gamma (\Sigma_a^{} \rightarrow \phi^\ast_{}+ \phi^\ast_{}) }{\Gamma (\Sigma_a^{} \rightarrow \psi^c_L + l^{c}_L)+\Gamma(\Sigma_a^{} \rightarrow \phi^\ast_{}+\phi^\ast_{})}\neq 0\,.
\end{eqnarray}
This CP asymmetry can arrive at a nonzero value if the models (\ref{lag5a}) and (\ref{lag5b}) contain at least two Higgs triplets $\Sigma_{1,2,...}^{}$, see Fig. \ref{sdecay}. After the Higgs triplets $\Sigma_a^{}$ go out of equilibrium, their CP-violating decays (\ref{cpa}) can generate an asymmetry stored in the vector-like lepton doublets $(\psi^{}_L,\psi'^{}_L)$ and the SM lepton doublets $l_L^{}$. The $\psi'^{}_L$ or $\psi^{}_L$ asymmetry eventually can contribute to the $l_L^{}$ asymmetry because of the $M_\psi^{}$ and $m_D^{}$ terms in Eqs. (\ref{lag5a}) and (\ref{lag5b}). The $l_L^{}$ asymmetry can be partially converted to a baryon asymmetry through the sphaleron \cite{krs1985} processes. This leptogenesis \cite{fy1986} scenario is very similar to the leptogenesis in type-II seesaw \cite{mz1992,ms1998}. The details can be found in literatures such as \cite{ms1998,hrs2006,adh2014}. Similarly, the other models (\ref{lag4a}-\ref{lag4b}) and (\ref{lag6a}-\ref{lag6b}) can accommodate a leptogenesis through the decays of the heavy fermion singlets or triplets.

In this work we have clarified the Weinberg dimension-5 operator can be induced at tree level by the vector-like lepton doublets in association with the unusual fermion singlets, Higgs triplets or fermion triplets, besides the well-known type-I, type-II and type-III seesaw. The vector-like lepton doublets can lead to rich observable phenomena if they are at the TeV scale. Meanwhile, when the fermion singlets, Higgs triplets or fermion triplets are very heavy, their out-of-equilibrium and CP-violating decays can accommodate a leptogenesis to explain the baryon asymmetry in the universe. Our scenario can be naturally embedded into some $SU(5)$ or $SO(10)$ GUTs without the conventional seesaw. In these $SU(5)$ or $SO(10)$ GUTs, the 5-dimensional or 10-dimensional representations for the vector-like lepton doublets can also result in the KSVZ invisible axion by their color-triplet components if a PQ symmetry is introduced properly.

\textbf{Acknowledgement}: I thank Hong-Jian He, Xiao-Gang He, Ernest Ma, Rabi N. Mohapatra and Utpal Sarkar for helpful comments and suggestions. This work was supported in part by the National Natural Science Foundation of China under Grant No. 11675100 and in part by the Fundamental Research Funds for the Central Universities.


\begin{thebibliography}{99}



 
\bibitem{pdg2018}
M. Tanabashi {\it et al.}, (Particle Data Group), Phys. Rev. D \textbf{98}, 030001 (2018).





\bibitem{weinberg1979}
S. Weinberg, Phys. Rev. Lett. \textbf{43}, 1566 (1979).


\bibitem{minkowski1977}
P. Minkowski, Phys. Lett. B \textbf{67}, 421 (1977).

\bibitem{yanagida1979}
T. Yanagida, {\it Proceedings of the Workshop on Unified Theory and the Baryon Number of the Universe}, ed. O. Sawada and A. Sugamoto (Tsukuba 1979).

\bibitem{grs1979}
M. Gell-Mann, P. Ramond, and R. Slansky, {\it Supergravity}, ed. F. van Nieuwenhuizen and D. Freedman
(North Holland 1979).

\bibitem{ms1980}
R.N. Mohapatra and G. Senjanovi\'{c}, Phys. Rev. Lett. \textbf{44}, 912 (1980).






\bibitem{mw1980}
M. Magg and C. Wetterich, Phys. Lett. B \textbf{94}, 61 (1980).

\bibitem{sv1980}
J. Schechter and J.W.F. Valle, Phys. Rev. D \textbf{22}, 2227 (1980).


\bibitem{cl1980}

T.P. Cheng and L.F. Li, Phys. Rev. D \textbf{22}, 2860 (1980). 

\bibitem{lsw1981}
G. Lazarides, Q. Shafi, and C. Wetterich, Nucl. Phys. B \textbf{181},
287 (1981).

\bibitem{ms1981}
R.N. Mohapatra and G. Senjanovi\'{c}, Phys. Rev. D
\textbf{23}, 165 (1981).



\bibitem{flhj1989}
R. Foot, H. Lew, X.G. He, and G.C. Joshi, Z. Phys. C \textbf{44},
441 (1989).

\bibitem{ma1998}
E. Ma, Phys. Rev. Lett. \textbf{81}, 1171 (1998).




\bibitem{mv1986}
R. Mohapatra and J. W. F. Valle, Phys. Rev. D \textbf{34}, 1642 (1986).




\bibitem{abp2008}
F. del Aguila, J. de Blas, M. Perez-Victoria, Phys. Rev. D \textbf{78}, 013010 (2008).




\bibitem{alsv1995}

E.K. Akhmedov, M. Lindner, E. Schnapka, and J.W.F. Valle, Phys. Rev. D \textbf{53}, 2752 (1996).




\bibitem{barr2004}
S.M. Barr, Phys. Rev. Lett. \textbf{92}, 101601 (2004). 


\bibitem{mrv2005}
 
M. Malinsky, J.C. Romao, and J.W.F. Valle, Phys. Rev. Lett. \textbf{95}, 161801 (2005).


\bibitem{ma2009}
E. Ma, Mod. Phys. Lett. A \textbf{24}, 2491 (2009).





\bibitem{ma2018}
E. Ma, Phys. Rev. D \textbf{98}, 091701 (2018).







\bibitem{pq1977}
R.D. Peccei and H.R. Quinn, Phys. Rev. Lett. \textbf{38}, 1440
(1977).



\bibitem{kim1979}
J.E. Kim, Phys. Rev. Lett. \textbf{43}, 103 (1979).



\bibitem{svz1980}
M.A. Shifman, A.I. Vainshtein, and V.I. Zakharov, Nucl. Phys. B \textbf{166}, 493
(1980).




\bibitem{krs1985}
V.A. Kuzmin, V.A. Rubakov, and M.E. Shaposhnikov, Phys. Lett. B
\textbf{155}, 36 (1985).




\bibitem{fy1986}
M. Fukugita and T. Yanagida, Phys. Lett. B \textbf{174}, 45 (1986).





\bibitem{mz1992}
R.N. Mohapatra and X. Zhang, Phys. Rev. D \textbf{46}, 5331 (1992).




\bibitem{ms1998}
E. Ma and U. Sarkar, Phys. Rev. Lett. \textbf{80}, 5716 (1998).

\bibitem{hrs2006}
T. Hambye, M. Raidal, and A. Strumia, Phys. Lett. B \textbf{632}, 667 (2006).



\bibitem{adh2014}
D. Aristizabal Sierra, M. Dhen, and T. Hambye, JCAP \textbf{1408}, 003 (2014).




\end{thebibliography}
\end{document}